\documentclass[12pt,preprint,flushrt]{aastex}
\usepackage{apjfonts}
\begin{document}

\title{On Competing Models of Coronal Heating and Solar Wind
Acceleration: \\
The Debate in '08}

\author{Steven R. Cranmer}
\affil{Harvard-Smithsonian Center for Astrophysics,
60 Garden Street, Cambridge, MA 02138}

\begin{abstract}
In preparation for lively debate at the May 2008 SPD/AGU Meeting
in Fort Lauderdale, this document attempts to briefly lay out my
own view of the evolving controversy over how the solar wind is
accelerated.
It is still unknown to what extent the solar wind is fed by flux
tubes that remain open (and are energized by footpoint-driven
wavelike fluctuations), and to what extent much of the mass and
energy is input more intermittently from closed loops into the
open-field regions.
It may turn out that a combination of the two ideas is needed
to explain the full range of observed solar wind phenomena.
\end{abstract}

\section{Introduction}

The solar corona is the hot, ionized outer atmosphere of
the Sun that expands into interplanetary space as a
supersonic solar wind (Parker 1958, 1965, 1991, 2001).
This tenuous and unbounded medium is a unique laboratory
for the study of magnetohydrodynamics (MHD) and plasma physics
with ranges of parameters (e.g., densities and temperatures)
that are inaccessible on Earth.
Despite more than a half-century of study, though, the basic
physical processes responsible for heating the million-degree
corona and accelerating the solar wind are still not known.
Identification of these processes is important not only for
understanding the origins and impacts of space weather
(e.g., Feynman \& Gabriel 2000; Cole 2003),
but also for establishing a baseline of knowledge about a
well-resolved star that is directly relevant to other
astrophysical systems.

Different physical mechanisms for heating the corona probably
govern active regions, closed loops in the quiet corona, and
the open field lines that give rise to the solar wind
(see reviews by Marsch 1999; Hollweg \& Isenberg 2002;
Longcope 2004; Gudiksen 2005; Aschwanden 2006; Klimchuk 2006).
The ultimate source of the energy is the solar convection zone
(e.g., Abramenko et al.\  2006; McIntosh et al.\  2007).
A key aspect of solving the ``coronal heating problem'' is thus
to determine how a small fraction of that mechanical energy is
transformed into magnetic and thermal energy above the photosphere.
It seems increasingly clear that loops in the low corona are
heated by small-scale, intermittent magnetic reconnection that
is driven by the continual stressing of their magnetic footpoints.
However, the extent to which this kind of impulsive energy
addition influences the acceleration of the solar wind is not
yet known.
Figure 1 illustrates a range of ideas concerning how the magnetic
topology varies between the coronal base and open flux tubes.

\begin{figure}[t]
\epsscale{0.99}
\plotone{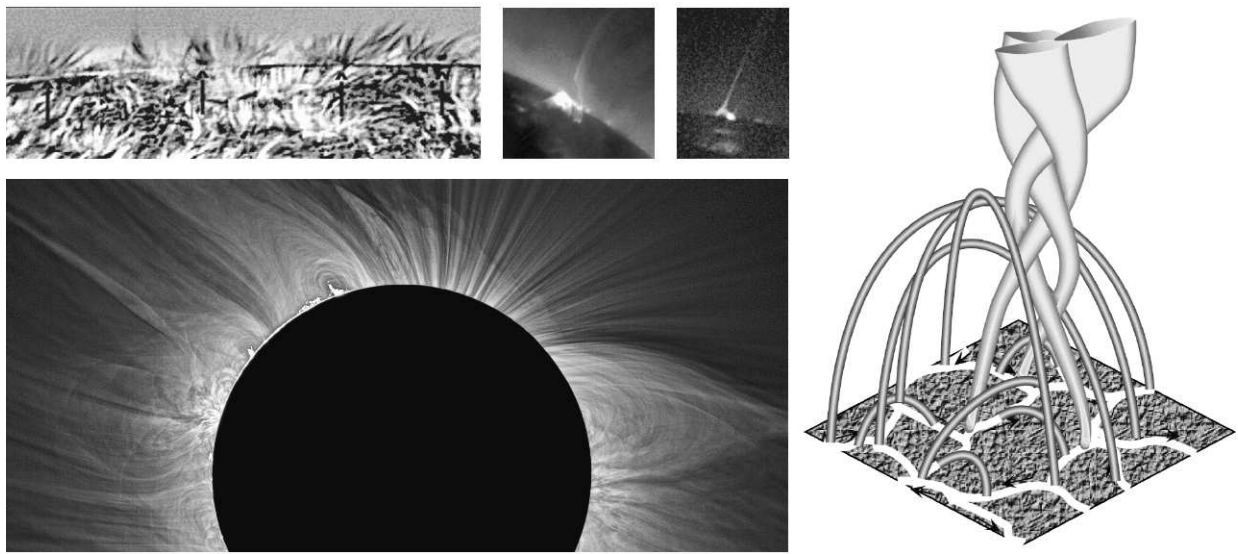}
\caption{Open and closed flux tubes juxtaposed throughout the corona.
{\em Clockwise from upper-left:}
image-processed H$\alpha$ polar chromospheric structures
(Filippov et al.\  2007);
two examples of X-ray jets from {\em Hinode}/XRT (e.g.,
Shimojo et al.\  2007);
illustration of interpenetrating supergranular loops and open
``canopy'' flux tubes (Fisk \& Zurbuchen 2006);
image-processed white-light corona from the 2007 March 29 eclipse
(Pasachoff et al.\  2007).}
\end{figure}

Intertwined with the coronal heating problem is the heliophysical
goal of being able to make accurate predictions of
how both fast and slow solar wind streams are accelerated.
Empirical correlation techniques have become more sophisticated
and predictively powerful (e.g., Wang \& Sheeley 1990, 2006;
Arge \& Pizzo 2000; Leamon \& McIntosh 2007; Cohen et al.\  2007;
Vr\v{s}nak et al.\  2007) but they are limited because they do
not identify or utilize the physical processes actually responsible
for solar wind acceleration.
There seem to be two broad classes of physics-based models that attempt
to self-consistently answer the question:
{\em ``How are fast and slow wind streams heated and accelerated?''}

\vspace*{0.03in}
\noindent
\newcounter{bean}
\begin{list}{\arabic{bean}.}{\usecounter{bean}%
\setlength{\leftmargin}{0.17in}%
\setlength{\rightmargin}{0.0in}%
\setlength{\labelwidth}{0.17in}%
\setlength{\labelsep}{0.05in}%
\setlength{\listparindent}{0.0in}%
\setlength{\itemsep}{0.05in}%
\setlength{\parsep}{0.0in}%
\setlength{\topsep}{0.0in}}

\item[1.]
In {\bf Wave/Turbulence-Driven (WTD) models,}
it is generally assumed that the convection-driven jostling of magnetic
flux tubes in the photosphere drives wave-like fluctuations that
propagate up into the extended corona.
These waves (usually Alfv\'{e}n waves) are often proposed to
partially reflect back down toward the Sun, develop into strong
MHD turbulence, and/or dissipate over a range of heights.
These models also tend to explain the differences between fast and
slow solar wind {\em not} by any major differences in the lower
boundary conditions, but instead as an outcome of different rates of
lateral flux-tube expansion over several solar radii ($R_{\odot}$)
as the wind accelerates
(see, e.g., Hollweg 1986; Wang \& Sheeley 1991; Matthaeus et al.\  1999;
Cranmer 2005; Suzuki 2006; Suzuki \& Inutsuka 2006;
Verdini \& Velli 2007; Cranmer et al.\ 2007).

\item[2.]
In {\bf Reconnection/Loop-Opening (RLO) models,}
the flux tubes feeding the solar wind are assumed to be influenced
by impulsive bursts of mass, momentum, and energy addition in the
lower atmosphere.
This energy is usually assumed to come from magnetic reconnection
between closed, loop-like magnetic flux systems (that are in the
process of emerging, fragmenting, and being otherwise jostled by
convection) and the open flux tubes that connect to the solar wind.
These models tend to explain the differences between fast and slow
solar wind as a result of qualitatively different rates of flux
emergence, reconnection, and coronal heating at the basal footpoints
of different regions on the Sun
(see, e.g., Axford \& McKenzie 1992, 2002; Fisk et al.\  1999;
Ryutova et al.\  2001;
Markovskii \& Hollweg 2002, 2004; Fisk 2003; Schwadron \& McComas 2003;
Woo et al.\  2004; Fisk \& Zurbuchen 2006).

\end{list}

\noindent
It is notable that both the WTD and RLO models have recently passed
some basic ``tests'' of comparison with observations.
Both kinds of model have been shown to be able to produce
fast ($v \gtrsim 700$ km/s), low-density wind from coronal holes and
slow ($v \lesssim 400$ km/s), high-density wind from streamers rooted
in quiet regions.
Both kinds of model also seem able to reproduce the observed
{\em in~situ} trends of how frozen-in charge states and the FIP effect
vary between fast and slow wind streams.

The fact that both sets of ideas described above seem to mutually
succeed at explaining the fast/slow solar wind {\em could}
imply that a combination of both ideas would work best.
However, it may also imply that the existing models do not yet
contain the full range of physical processes---and that once these
are included, one or the other may perform noticeably better than
the other.
It also may imply that the comparisons with observations have
not yet been comprehensive enough to allow the true differences
between the WTD and RLO ideas to be revealed.

Several recent observations have pointed to the importance of
understanding the relationships and distinctions between the
WTD and RLO models.
{\em Hinode} has shown that open-field regions are filled with
numerous coronal jets that may contribute a sizable fraction
of the solar wind mass flux
(Culhane et al.\  2007; Cirtain et al.\  2007;
Shimojo et al.\  2007).
Also, new observations of Alfv\'{e}n waves above the solar limb
indicate the highly intermittent nature of how kinetic energy is
distributed in spicules, loops, and the open-field corona
(Tomczyk et al.\  2007; De Pontieu et al.\  2007).
Spectroscopic observations of blueshifts in the chromospheric
network have long been interpreted as the launching points of
solar wind streams, but it remains unclear how nanoflare-like
events or loop-openings contribute to these diagnostics
(He et al.\  2007; Aschwanden et al.\  2007; McIntosh et al.\  2007).
Even out in the {\em in~situ} solar wind---far above the roiling
``furnace'' of flux emergence at the Sun---there remains evidence
for ongoing reconnection (Gosling et al.\  2005; Gosling 2007)
as well as turbulence timescales that may come from flux
cancellation in the low corona (Hollweg 1990, 2006).

Determining whether the WTD or RLO paradigm---or some combination
of the two---is the dominant cause of global solar wind variability
is a key prerequisite to building physically realistic predictive
models of the heliosphere (Zurbuchen 2006, 2007).
Many of the widely-applied global modeling codes (e.g.,
Riley et al.\  2001; Roussev et al.\  2003; T\'{o}th et al.\  2005;
Usmanov \& Goldstein 2006; Feng et al.\  2007)
continue to utilize relatively simple empirical prescriptions
for coronal heating in the energy conservation equation.
Improving the identification and characterization of the
key physical processes will provide a clear pathway for
inserting more physically realistic coronal heating ``modules''
into 3D MHD codes.

This paper does not go into detail about the specific
avenues of future work, both theoretical and observational, that
need to be pursued in order to better determine the relative
contributions of the WTD and RLO processes.
Thus, the following two sections just give some additional
information about the WTD and RLO paradigms as they stand in 2008.

\section{The Wave/Turbulence-Driven (WTD) Solar Wind Idea}

There has been substantial work over the past few decades put into
exploring the idea that coronal heating and solar wind acceleration
along open flux tubes may be explained as a result of wave dissipation
and turbulent cascade.
No matter the relative importance of reconnections and
loop-openings in the low corona, we do know that waves and turbulent
motions are present everywhere from the photosphere to the heliosphere,
and it is important to determine how they affect the mean state
of the plasma.
Although this section mainly describes recent work by the author,
these results would not have been possible without the earlier work
on wave/turbulent heating by, e.g.,
Coleman (1968), Hollweg (1986), Isenberg (1990), Li et al.\  (1999),
Matthaeus et al.\  (1999), Dmitruk et al.\  (2001, 2002) and
many others.

Cranmer et al.\  (2007) described a set of models that solve for
the time-steady plasma properties
along a 1D magnetic flux tube that is rooted in the
solar photosphere and expands into interplanetary space.
The numerical code developed in this work, called {\sc zephyr,}
solves the one-fluid equations of mass, momentum,
and energy conservation simultaneously with transport
equations for Alfv\'{e}nic and acoustic wave energy.
{\sc zephyr} is the first code capable of producing self-consistent
solutions for the photosphere, chromosphere, corona, and solar wind
that combine:
(1) shock heating driven by an empirically
guided acoustic wave spectrum, (2) extended heating from
Alfv\'{e}n waves that have been partially reflected, then damped
by anisotropic turbulent cascade, and (3) wind acceleration
from gradients of gas pressure, acoustic wave pressure, and
Alfv\'{e}n wave pressure.

The only input ``free parameters'' to {\sc zephyr} are the photospheric
lower boundary conditions for the waves and the radial dependence
of the background magnetic field along the flux tube.
The majority of heating in these models comes from the
turbulent dissipation of partially reflected Alfv\'{e}n waves
(see also Matthaeus et al.\  1999; Dmitruk et al.\  2002;
Verdini \& Velli 2007).
Measurements of G-band bright points in the photosphere were used
to constrain the Alfv\'{e}n wave power spectrum at the lower boundary,
and non-WKB wave transport equations were solved to determine
the degree of linear reflection.
The resulting values of the Elsasser amplitudes $Z_{\pm}$, which
denote the energy contained in upward ($Z_{-}$) and downward
($Z_{+}$) propagating waves, were then used to constrain the
energy flux in the cascade.
Cranmer et al.\  (2007) used a phenomenological form for the damping
rate that has evolved from studies of Reduced MHD and
comparisons with numerical simulations.
The resulting heating rate (erg s$^{-1}$ cm$^{-3}$) is given by
\begin{displaymath}
  Q \, = \, \rho \, \left(
  \frac{1}{1 + [ t_{\rm eddy} / t_{\rm ref} ]^{n}} \right) \,
  \frac{Z_{-}^{2} Z_{+} + Z_{+}^{2} Z_{-}}{4 L_{\perp}}
\end{displaymath}
(e.g., Hossain et al.\  1995; Zhou \& Matthaeus 1990).
The transverse length scale $L_{\perp}$ is an effective
perpendicular correlation length of the turbulence, and
Cranmer et al.\  (2007) used
a standard assumption that $L_{\perp}$ scales with the
cross-sectional width of the flux tube (Hollweg 1986).
The term in parentheses above is an efficiency factor that
accounts for situations in which the cascade does not have time to
develop before the waves or the wind carry away the energy
(Dmitruk \& Matthaeus 2003).
The cascade is ``quenched''
when the nonlinear eddy time scale $t_{\rm eddy}$ becomes much
longer than the macroscopic wave reflection time scale $t_{\rm ref}$.
In most of the models, Cranmer et al.\  (2007)
used $n=1$ based on analytic and numerical
models (Dobrowolny et al.\  1980; Oughton et al.\  2006),
but they also tried $n=2$ to explore a stronger form of this quenching.

\begin{figure}[t]
\epsscale{0.99}
\plotone{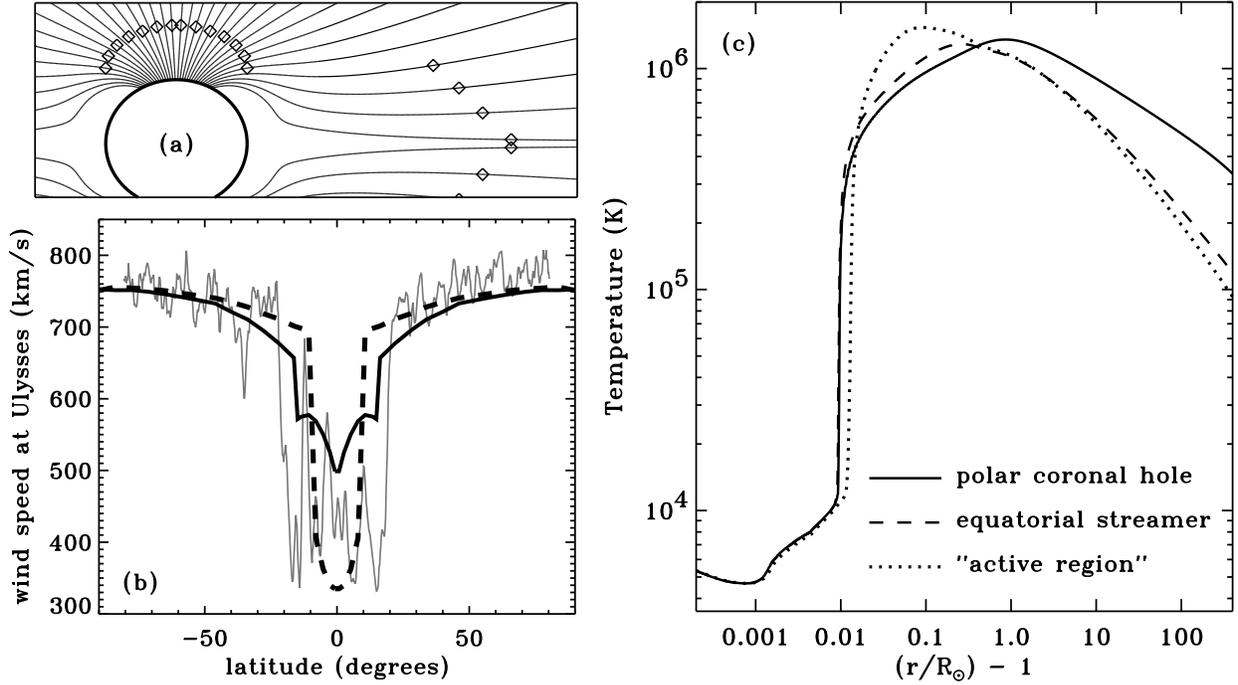}
\caption{Summary of recent WTD models.
{\em (a)} The adopted solar-minimum field geometry
(Banaszkiewicz et al.\  1998)
with radii of wave-modified critical points marked by symbols.
{\em (b)} Latitudinal dependence of wind speed at $\sim$2 AU
for models with $n=1$ (thick black curve) and $n=2$ (dashed curve),
compared with data from the first {\em Ulysses} polar pass in
1994--1995 (thin gray curve; Goldstein et al.\  1996).
{\em (c)} $T(r)$ for coronal hole
(solid curve), streamer-edge (dashed curve), and
strong-field active region (dotted curve) models.
Further details can be found in Cranmer et al.\  (2007).}
\end{figure}

Figure 2 summarizes the results of varying the magnetic field
properties while keeping the lower boundary conditions fixed.
For a single choice for the photospheric wave properties,
the models produce a realistic range of slow and fast solar wind
conditions.
Specifically, a 2D model of coronal holes and
streamers at solar minimum reproduces the latitudinal
bifurcation of slow and fast streams seen by {\em Ulysses.}
The radial gradient of the Alfv\'{e}n speed affects where the
waves are reflected and damped, and thus whether energy
is deposited below or above the Parker critical point.
As predicted by earlier studies, a larger coronal ``expansion
factor'' gives rise to a slower and denser wind, higher temperature
at the coronal base, and less intense Alfv\'{e}n waves at 1 AU.

Perhaps more surprisingly, varying the coronal expansion factor also
produces correlative trends that are in good agreement with {\em
in~situ} measurements
of commonly measured ion charge state ratios (e.g.,
O$^{7+}$/O$^{6+}$) and FIP-sensitive abundance ratios (e.g., Fe/O).
Cranmer et al.\  (2007) showed that the slowest solar wind
streams---associated with active-region fields at the base---can
produce a factor of $\sim$30 larger frozen-in ionization-state ratio of
O$^{7+}$/O$^{6+}$ than high-speed streams from polar coronal holes,
despite the fact that the temperature at 1~AU is lower in
slow streams than in fast streams.
Also, when elemental fractionation is modeled using Laming's (2004)
theory of preferential wave-pressure acceleration, the slow wind
streams exhibit a substantial relative buildup of elements with low
First Ionization Potential (FIP) with respect to the
high-speed streams.
Although the WTD models utilize identical photospheric lower boundary
conditions for all of the flux tubes, the self-consistent solutions
for the upper chromosphere, transition region, and low corona are
qualitatively different.
Feedback from larger heights (i.e., from variations in the
flux tube expansion rate and the resulting heating rate) extends
downward to create these differences.

\section{The Reconnection/Loop-Opening (RLO) Solar Wind Idea}

It is clear from observations of the Sun's ``magnetic carpet''
(Schrijver et al.\  1997; Title \& Schrijver 1998; Hagenaar et al.\  1999)
that much of coronal heating is driven by the continuous interplay
between the emergence, separation, merging, and cancellation
of small-scale magnetic elements.
Reconnection seems to be the most likely channel for the injected magnetic
energy to be converted to heat (e.g., Priest \& Forbes 2000).
Only a small fraction of the photospheric magnetic flux is in
the form of {\em open} flux tubes connected to the heliosphere
(Close et al.\  2003).
Thus, the idea has arisen that the dominant source of energy for open
flux tubes is a series of stochastic reconnection events between
the open and closed fields (e.g., Fisk et al.\  1999;
Ryutova et al.\  2001; Fisk 2003;
Schwadron \& McComas 2003; Schwadron et al.\  2006).

The natural appeal of the RLO idea is evident from the fact that
open flux tubes are always rooted in the vicinity of closed loops
(e.g., Dowdy et al.\  1986) and that all layers of the solar atmosphere
seem to be in continual motion with a wide range of timescales.
Indeed, observed correlations between the lengths of closed loops
in various regions, the electron temperature in the low corona, and
the wind speed at 1 AU (Feldman et al.\  1999;
Gloeckler et al.\  2003) are highly suggestive of a net transfer
of Poynting flux from the loops to the open-field regions that may
be key to understanding the macroscopic structure of the solar wind.
The proposed RLO reconnection events may also be useful in
generating energetic particles and cross-field diffusive transport
throughout the heliosphere (e.g., Fisk \& Schwadron 2001).

Testing the RLO idea using theoretical models seems to be more difficult
than testing the WTD idea because of the complex multi-scale nature
of magnetic reconnection.
It can be argued that one needs to create fully 3D models of the
coronal magnetic field (arising from multiple magnetic elements on
the surface) to truly assess the full range of
closed/open flux interactions.
The idea of modeling the coronal field via a collection of
discrete magnetic sources (referred to in various contexts as
``magneto-chemistry,'' ``tectonics,'' or ``magnetic charge topology'')
has been used extensively to study the evolution of
the closed-field corona
(e.g., Longcope 1996; Schrijver et al.\  1997;
Longcope \& Kankelborg 1999; Sturrock et al.\  1999;
Priest et al.\  2002; Beveridge et al.\  2003;
Barnes et al.\  2005; Parnell 2007)
but applications to open fields and the solar wind have been rarer
(see, however, Fisk 2005; Tu et al.\ 2005).

\end{document}